\documentclass{elsart} \usepackage{epsfig}
\renewcommand{\i}{\ensuremath{\mathrm{i}}}
\newenvironment{minifig}[1][b]{\begin{minipage}[#1]{.46\linewidth}}
  {\end{minipage}}
\begin{document}
%
\begin{frontmatter}
  \title{Vector meson radiation in relativistic heavy-ion collisions}
  \author{Bryan E. Barmore\thanksref{email}} \thanks[email]{email:
    barmore@physics.wm.edu} \address{Department of Physics, College of
    William and Mary, Williamsburg, VA 23185}
\begin{abstract}
  The ($\sigma, \omega$) model in mean-field approximation where the
  meson fields are treated classically, describes much of observed
  nuclear structure and has been employed to describe the nuclear
  equation of state up to the quark-gluon phase transition.  The
  acceleration of the meson sources, for example, in relativistic
  heavy-ion collisions, should result in bremsstrahlung-like radiation
  of the meson fields.  The many mesons emitted serve to justify the
  use of classical meson fields.  The slowing of the nuclei during the
  collision is modeled here as a smooth transition from initial to
  final velocity.  Under ultra-relativistic conditions, vector
  radiation dominates.  The angular distribution of energy flux shows
  a characteristic shape.  It appears that if the vector meson field
  couples to the conserved baryon current, independent of the baryonic
  degrees of freedom, this mechanism will contribute to the radiation
  seen in relativistic heavy-ion collisions.  The possible influence
  of the quark-gluon plasma is also considered.
\end{abstract}
\end{frontmatter}

\section{Introduction}

The $(\sigma, \omega)$ model of the nucleus~\cite{W95,SW86} is a
relativistic quantum field theory which describes the nuclear
interaction using three fields.  They are neutral scalar meson,
neutral vector meson and baryon fields.  The scalar meson field
couples to the scalar density, $\bar{\psi} \psi$, while the vector
meson field couples to the conserved baryon current, $B_{\mu} \equiv
\bar{\psi} \gamma_{\mu} \psi$.  In the relativistic mean-field
approximation (RMFT) where the sources are large, the meson fields can
be replaced by classical fields and the linearized Dirac equation
solved exactly.  This model has had several successes, properly
describing: bulk properties of nuclear matter, ground state properties
of nuclei, the excitation spectrum of nuclei, low energy
nucleon-nucleus scattering observables, as well as collective motion.
All except the first of the preceding list incorporate spatial
dependence for the meson fields and their sources.  The last item,
collective motion, incorporates a slow time dependence in addition to
the spatial dependence.

It is here proposed to use this model to calculate vector meson
production during relativistic heavy-ion collisions like those to be
seen at RHIC\@.  All of the above successes of the $(\sigma, \omega)$
model have been for low and intermediate energy phenomena.  Therefore,
some justification is necessary for use of this model in a radically
different physical situation.  First, using ground state
wave-functions calculated from this model and empirical
nucleon-nucleon scattering amplitudes one can obtain excellent
agreement with experimental nucleon-nucleus scattering observables in
the relativistic impulse approximation up to energies of the order 1
GeV~\cite{Se87}; second, the longitudinal response for
${}^{56}$Fe($e$,$e'$) at momentum transfer of $|\vec{q}| = 0.55$ GeV
has been calculated by Frank~\cite{Fr94} in this model and agrees well
with data; in addition, the $(\sigma, \omega)$ model taken in RMFT has
been employed to describe the nuclear equation of state up to the
quark-gluon phase transition~\cite{W95,Ch78}.  Furthermore, the
presence of many quanta of the meson fields here serves to validate
the classical approximation.

It is not clear what the appropriate degrees of freedom actually are
for the RHIC-like collisions.\footnote{In the work of Mishustin
  \emph{et. al}~\protect\cite{MS95} it is stated, ``One cannot even
  say what degrees of freedom, hadrons or quarks and gluons, are more
  suitable for describing these collisions.''} Although the ($\sigma,
\omega$) model deals with hadrons, these results will turn out to be
more general.  We consider a process whereby two conserved baryon
currents pass through each other, are slowed, and thus radiate energy
via bremsstrahlung in the form of vector mesons.  In the end, the only
requirements for this calculation are an empirical knowledge of the
initial and final rapidities of the baryon currents, conservation of
these currents, and a vector meson that couples to the conserved
baryon currents.  In the final analysis, it is irrelevant whether the
baryon current is carried by nucleons or quarks during the collision,
as long as the vector meson couples to the baryon current.

It should be noted that in the work of Mishustin \textit{et.\ al}\ 
\cite{MS95}, where the $(\sigma, \omega)$ model is employed to
calculate baryon--anti-baryon production via virtual vector meson
bremsstrahlung under RHIC conditions, it is suggested that real
$\omega$-mesons should be produced via bremsstrahlung and
that this process should be relatively soft.\footnote{%
  In Ref.~\cite{MS95} it is further observed about vector meson
  bremsstrahlung that ``... it is clear that the same mechanism can
  produce also mesons.  For instance, the real $\omega$-meson can be
  generated in the bremsstrahlung process when the four-momenta of
  quanta satisfy the mass shell constraint $p^{2} = m_{\omega}^{2}$.
  ... These channels are characterized by lower threshold and,
  therefore, by smaller momentum transfers.  Since the corresponding
  coupling constants are also large, one can expect high
  multiplicities of mesons coherently produced in ultra-relativistic
  nuclear collisions.''}

In this work only central collisions of identical heavy nuclei are
considered.  Therefore, the lab and equal-velocity frames are
equivalent.  In Ref.~\cite{H94} Hanson shows evidence that for central
high-energy collisions (rapidity larger than 3), the nuclei are
expected to be slowed by the collision, but not completely stopped.
This slowing occurs over the short time of the collision.  This gives
rise to a large deceleration.  In this model, the decelerating sources
radiate the classical meson fields.  This is the same process as
classical electromagnetic bremsstrahlung for an accelerated charge.

Due to Lorentz contraction, the scalar meson radiation is greatly
suppressed relative to the vector meson radiation.  Most of the vector
meson radiation is in the form of high energy mesons whose mass is
therefore neglected.  An attenuation factor is included to account for
the strong interaction between the vector mesons and the baryons.  The
attenuation factor is constructed in a Lorentz invariant manner to
maintain the covariance of the theory.  This model contains three
adjustable parameters: the deceleration time, the rapidity loss and
the total cross section for the vector meson-nucleon interaction.  The
angular distribution of energy flux is characteristic, varying only in
magnitude for variations in the cross section and deceleration time.
Variations in rapidity loss affect both the magnitude and width of the
angular distribution; however, the general shape is robust for
parameter variations.

This work indicates that the bremsstrahlung of vector mesons could
contribute significantly to the total radiated energy during
relativistic heavy-ion collisions.  The model used here is very
simple; it depends only on the vector mesons coupling to a conserved
baryon current.  The fact that this model predicts an appreciable
amount of radiation with a characteristic angular distribution that is
robust against parameter variation suggests that it is deserving of
further investigation.

This radiation of mesons during relativistic heavy-ion collisions has
been considered before.  Weber \emph{et.\ al}~\cite{WB90} examined the
dynamics of relativistic heavy-ion collisions using the ($\sigma,
\omega$) model in mean field theory.  They modeled the collisions
within the relativistic Boltzmann-Uehling-Uhlenbeck model using full
solutions to the meson-field equations.  At the energies they studied
(1-20~GeV/A), they found meson radiation to be negligible ($<$ 3
MeV/nucleon).  However, the amount of energy radiated increases as the
acceleration squared, so their work does not rule out appreciable
radiation at higher energies.  The production of pions and photons via
bremsstrahlung was investigated in Refs. \cite{VMG80,SU87}.
Ivanov~\cite{I89} calculated $\omega$-meson radiation due to
filamentation instability.  The behavior of the meson fields during
relativistic heavy-ion collisions was explored in
Refs.~\cite{Cu85,Ba87}; however, these are at much lower bombarding
energies and did not directly look at meson radiation.

It is expected that at the energies available at RHIC these central
collisions will produce a quark-gluon plasma.  The question then
arises, what effect will the creation of the plasma have on this model
that uses baryons and mesons?  Indeed, would any sign of this hadronic
process survive under these conditions?  It has been suggested that as
individual quarks interact during the collision, color strings or flux
tubes will form~\cite{Bi82,Bi85,Ka89}.  These flux tubes are stretched
out behind the nuclei as they pass through each other (see, for
example, figure 4 of Ref.~\cite{Bi82}).  The breaking of these flux
tubes then provides an important contribution to the formation of the
quark-gluon plasma.  The meson production described in the present
work comes exclusively from the forward going baryons that constitute
the initial colliding nuclei.  Even if the mesons must pass through
part of the baryon-rich quark-gluon plasma, there is theoretical
evidence that meson-like modes can propagate through a quark-gluon
plasma~\cite{Ha85}.  Such a physical situation may alter the
attenuation but could still leave a detectable sign of this mechanism.

This paper attempts to test the limits of the $(\sigma, \omega)$ model
in RMFT for a strong time-dependence.  Section~\ref{sec:formal}
introduces the framework for the $(\sigma, \omega)$ model and
classical bremsstrahlung.  Sections~\ref{sec:current} to
\ref{sec:attenuation} discuss the model for the baryon current, the
degree of incoherence in the radiation and the attenuation of the
vector meson radiation.  The energy spectrum is investigated in
section~\ref{sec:massless}.  Sections~\ref{sec:results} and
\ref{sec:conclusions} discuss the results and conclusions.

\section{Formalism}
\label{sec:formal}

The basic $(\sigma, \omega)$ model is defined by the lagrangian
density,\footnote{Throughout this paper natural units are used,
  $c=\hbar=1$.  In these units 0.197 GeV = 1 fm$^{-1}$.  The metric is
  that of Bjorken and Drell and Ref.~\protect\cite{SW86}.}
\begin{eqnarray}
  {\mathcal L} & = & -\frac{1}{4}F^{\mu \nu} F_{\mu \nu} + \frac{1}{2}
  m_{\omega}^{2} V^{\mu} V_{\mu} + \frac{1}{2} \left[
    \left(\frac{\partial \phi}{\partial x_{\mu}} \right)^{2} -
    m_{\sigma}^{2} \phi^{2} \right] \nonumber \\ & & + \bar{\psi}
  \left[ \i \gamma_{\mu} (\frac{\partial}{\partial x_{\mu}} -
    g_{v}V^{\mu}) - (M_{N} - g_{s}\phi) \right] \psi ,
\end{eqnarray}
where $V_{\mu}$ is the vector field, $\phi$ is the scalar field and
$\psi$ is the baryon field.  The masses of the nucleon, vector meson
and scalar meson are $M_{N}$, $m_{\omega}$ and $m_{\sigma}$
respectively.  The field tensor is defined as
\begin{equation}
  F^{\mu \nu} = \frac{\partial V^{\nu}}{\partial x_{\mu}} -
  \frac{\partial V^{\mu}}{\partial x_{\nu}}.
\end{equation}
This lagrangian neglects non-linear self-couplings of the scalar
field.  In the nuclear ground state, the vector and scalar fields are
of comparable strength~\cite{W95}.  Under a Lorentz boost the scalar
density is invariant; however, the vector density is enhanced by a
factor of $\gamma$, defined as
\begin{equation}
\label{lorentz}
\gamma = (1 - \beta^{2})^{-1/2} ,
\end{equation}
with $\beta$ the velocity.  For the dynamics of the collisions
investigated in this paper, this factor varies between 10 and 100.
The radiated energy is proportional to the field squared; therefore,
the vector meson radiation is at least a factor of 100 greater than
the scalar meson radiation.  Consequently, \emph{the scalar field can
  be neglected}.

The energy-momentum tensor is,
\begin{equation} 
  T^{\mu \nu} = - {\mathcal L} g^{\mu \nu} + \frac{\partial {\mathcal
      L}} {\partial(\partial q / \partial x^{\mu})} \frac{\partial q}
  {\partial x_{\nu}} = \frac{1}{4} g^{\mu \nu} F^{\alpha \beta}
  F_{\alpha \beta} - F^{\mu}_{\mbox{ }\sigma} \frac{\partial
    V^{\sigma}}{\partial x_{\nu}}.
\end{equation}
This form of the energy-momentum tensor works well for calculating the
total energy and momentum of the system.  However, it contains total
divergences that will give incorrect results for the energy flux.
Therefore, the symmetric energy-momentum tensor must be used
\cite{Ad90},
\begin{equation}  \label{en-mom}
  \Theta^{\mu \nu} = \frac{1}{4} g^{\mu \nu} F^{\alpha \beta}
  F_{\alpha \beta} + F^{\mu \sigma} F_{\sigma}^{\mbox{ } \nu} .
\end{equation}

The power radiated\footnote{The development of the radiated energy
  follows closely that in Jackson~\protect\cite{JA75}.} is
\begin{equation}
   \label{power}
   \frac{\d P(t')}{\d \Omega} = R^{2} [\vec{S}(t) \cdot \vec{n}]_{{\rm
       ret}}
\end{equation}
where $P(t')$ is the power radiated at time $t'$ and $S_{i} (\equiv
\Theta^{0i})$ is the Poynting vector.  The observer is located a
distance $R$ from the source of the radiation along the unit vector
$\vec{n}$.  The notation, $[]_{\rm ret}$, means evaluate at $t' = t -
R(t')$.

The equation of motion for the vector meson field is
\begin{equation}
  \partial_{\nu} F^{\mu \nu} + m_{\omega}^{2} V^{\mu} = g_{v} B^{\mu}.
\end{equation}
Since the baryon current is conserved, this reduces to
$(\partial_{\nu}^{2} + m_{\omega}^{2}) V^{\mu} = g_{v} B^{\mu}$.  This
is the inhomogeneous Klein-Gordon equation for a massive particle
which has the solution,
\begin{equation}
  V^{\mu}(x) = \frac{1}{(2\pi)^{4}} \int \d ^{4} k \e^{-\i k \cdot x}
  \frac{g_{v} B^{\mu}(k)}{m_{\omega}^{2} - k^{2} - \i \eta
    \varepsilon}.
\end{equation}
In the above equation, $x$ is the four-position $(t,\vec{x})$, $k$ is
the four-momentum ($\varepsilon, \vec{k}$) and $B^{\mu}(k)$ is the
Fourier transform of the current.  A convergence factor, $\i \eta
\varepsilon$, is introduced to eliminate the singularity at $k^{2} =
m_{\omega}^{2}$.  The choice of $- \i \eta \varepsilon$ gives the
retarded solution.  The solution above can be converted to an integral
over $\d ^{4} x'$ and gives \mbox{$V^{\mu}(x) = \int \d ^{4} x'
  {\mathcal G}_{r}(x-x') B^{\mu}(x')$} where
\begin{eqnarray}
\label{green}
{\mathcal G}_{r}(x-x') & = & \frac{1}{(2\pi)^{4}} \int \d ^{4} k
\frac{\e^{-ik \cdot (x-x')}}{m_{\omega}^{2} - k^{2} - \i \eta
  \varepsilon} \nonumber \\ & = & \frac{\delta(\tau - {\mathcal
    R})}{4\pi {\mathcal R}} - \frac{m_{\omega}}{4\pi\sqrt{\tau^2 -
    {\mathcal R}^2}} J_{1} (m_{\omega}\sqrt{\tau^2 - {\mathcal R}^2})
\Theta(\tau - {\mathcal R}),
\end{eqnarray}
where $\tau = t - t'$ and ${\mathcal R} = |\vec{x} - \vec{x}'|$.
$J_{1}$ is the first order Bessel function and $\Theta(y)$ is the
Heavyside step function defined as $\Theta (y) \equiv (1 + |y|/y) /
2$.

If the baryon current is replaced by the electromagnetic current, the
coupling constant changed to the fine structure constant, $g_v^2/4\pi
\rightarrow \alpha$, and the mass, $m_{\omega}$, goes to zero, this
reduces to electromagnetic bremsstrahlung.  Note that as $m_{\omega}
\rightarrow 0$ the second term in equation~(\ref{green}) vanishes.  It
is argued in section~\ref{sec:massless} that neglecting the mass of
the vector meson does not alter the results appreciably, and we will
take $m_{\omega} = 0$ for the rest of this work.

The above solution for $V^{\mu}$ and equation~(\ref{en-mom}) gives a
distribution of energy as a function of outgoing energy,
$\varepsilon$, and angle of, (compare equation (14.70) of
Ref.~\cite{JA75})
\begin{equation}
  \label{freq}
  \frac{\d ^{2} E}{\d \varepsilon \, \d \Omega} =
  \frac{\varepsilon^{2} g_{v}^{2}}{4 \pi^{2}} \: \left| \;
    \int_{-\infty}^{\infty} \d t \int \d ^{3}x \, \vec{n} \times
    [\vec{n} \times \vec{B}(\vec{x}, t)] \, \e^{\i \varepsilon (t -
      \vec{n} \cdot \vec{x})} \; \right|^{2}.
\end{equation}
Again, $\vec{n}$ is the unit vector from the point of radiation to the
observer.

The radiation is strongly peaked in the direction of the baryon
cluster's motion.  Thus, we concentrate now on just one of the
incident nuclei.  Assuming the current is that of a collection of
point baryons traveling with identical velocity,
\begin{equation}
  \vec{B}(\vec{x},t) = \sum_j \delta(\vec{x} - \vec{x}_j(t))
  \vec{\beta}(t)
\end{equation}
we may do the spatial integration.  The baryon's position is given as
$\vec{x}_j(t) = \vec{X}(t) + \vec{r}_j(t)$ where $\vec{X}(t)$ is the
position of the center-of-mass and $\vec{r}_j(t)$ is the baryon's
position relative to the center-of-mass.  The relative position of the
baryon will vary with time, for example, due to changing Lorentz
contraction and the empirically seen spreading of final rapidity;
however, in this simple model this motion will be neglected relative
to the much more important center-of-mass motion.  The baryons will
this be considered as ``frozen'' relative to the baryon clusters as
they pass through each other and decelerate.  The baryon's position
thus becomes,
\begin{equation}
  \vec{x}_{j}(t) = \int_{-\infty}^{t} \vec{\beta}(t') \d t' +
  \vec{r}_{j}(-\infty) \equiv \vec{X}(t) + \vec{r}_{j}.
\end{equation}
Separating out the dependence on $j$ we may write,
\begin{equation}
\label{eq:freq3}
\frac{\d ^{2} E}{\d \varepsilon \, \d \Omega} = \frac{\varepsilon^{2}
  g_{v}^{2}}{4 \pi^{2}} \: \left| \; \int_{-\infty}^{\infty} \d t \,
  \vec{n} \times [\vec{n} \times \vec{\beta}(t)] \, \e^{\i \varepsilon
    (t - \vec{n} \cdot \vec{X}(t))} \; \sum_j \e^{-\i \varepsilon
    \vec{n} \cdot \vec{r}_j} \right|^{2}.
\end{equation}
Taking the sum outside the square gives a coherence factor
\begin{equation}
\label{eq:P}
{\mathcal P} \equiv \sum_{j=1}^{A} \sum_{k=1}^{A} \e^{-\i \varepsilon
  \, \vec{n} \cdot (\vec{r}_{j} - \vec{r}_{k})} .
\end{equation}
One must now take a statistical average over the positions of the
baryons in the incident nucleus; this average is weighted by the
square of the ground state wave-function.  If the exponent is large,
as argued in section~\ref{sec:incoherence}, the off-diagonal elements
average to zero and the coherence factor is approximately $A$, the
number of baryons.

We now wish to find the angular distribution of energy flux and the
total energy radiated.  Equation~(\ref{eq:freq3}) must therefore be
integrated over all frequencies.  Equation~(\ref{eq:freq3}) may be
rewritten as
\begin{equation}
  \frac{\d ^{2} E}{\d \varepsilon \, \d \Omega} =
  \frac{g_{v}^{2}}{4 \pi^{2}} \; A \left| \;
    \int_{-\infty}^{\infty} \d t \, \frac{\vec{n} \times [\vec{n}
      \times \vec{\dot{\beta}}(t)]}{(1 - \vec{n} \cdot
      \vec{\beta}(t))^2} \e^{\i \varepsilon (t - \vec{n} \cdot
      \vec{X}(t))} \right|^2.
\end{equation}
where we have assumed the velocity and acceleration are collinear to
write
\begin{equation}
  \frac{\d}{\d t} \left[ \frac{\vec{n} \times [\vec{n} \times
      \vec{\beta}(t)]}{1- \vec{n} \cdot \vec{\beta}(t)} \right] =
  \frac{\vec{n} \times [\vec{n} \times \vec{\dot{\beta}}(t)]} {(1-
    \vec{n} \cdot \vec{\beta}(t))^2}.
\end{equation}
Changing variables to $\tau = t - \vec{n} \cdot \vec{X}(t)$ and
expanding the square we may now do the $\varepsilon$ integration.  The
integration should be over positive energies; however, the integrand
is even in $\varepsilon$ so we take $1/2$ the integral over all
energies.  The angular distribution of energy flux is
\begin{equation}
\label{eq:angular}
\frac{\d E}{\d \Omega} = \frac{g_{v}^{2}}{4 \pi}\; A
\int_{-\infty}^{\infty} \d t \, \frac{\left| \vec{n} \times [\vec{n}
    \times \vec{\dot{\beta}}(t)] \right|^2} {(1- \vec{n} \cdot
  \vec{\beta}(t))^5}.
\end{equation}
The angular integration can also be done and this gives
\begin{equation}
  \label{eq:totalE}
  E = \frac{2 g_{v}^{2}}{3}\; A \int_{-\infty}^{\infty} \d t \,
  \gamma^6 |\vec{\dot{\beta}}(t)|^2.
\end{equation}
Note that the last two results are just the single particle result
(see equations (14.38) and (14.43) of Ref.~\cite{JA75}) times the
coherence factor, $\mathcal{P} \approx A$.

\section{Baryon Current Models}
\label{sec:current}

The only quantity required to evaluate equation (\ref{freq}) is the
baryon current, $\vec{B}(\vec{x},t)$.  In the present covariant
approach, this is related to the particles' velocity, $\vec{\beta}$,
in the lab frame by
\begin{equation}
  B_{\mu}(\vec{x}, t) = \sum_{\lambda = 1,2}
  \rho_B^{(\lambda)}(\vec{x}) U_{\mu}^{(\lambda)}(t)
\end{equation}
where 1 and 2 are the two nuclei, $\rho_B$ is the rest-frame baryon
density and $U_{\mu} = (\gamma, \gamma \vec{\beta})$ is the
four-velocity of the nucleus.  The baryon current is conserved;
\begin{equation}
  \frac{\partial}{\partial x_{\mu}} B_{\mu} = 0 ,
\end{equation}
and transforms as a four-vector, $ \Lambda_{\mu}^{\mbox{ }\nu}
B_{\nu}(\Lambda x) = \sum_{\lambda = 1,2} \rho_B^{(\lambda)}(\Lambda
x) \Lambda_{\mu}^{\mbox{ }\nu} U_{\nu}^{(\lambda)} (\Lambda x)$.

The baryon density should be determined from the ground-state nuclear
wave-functions, computed in the Dirac-Hartree approximation, summed
over the occupied orbitals.  This procedure is explained in detail in
Ref.~\cite{HS81} and implemented in Ref.~\cite{HS91}; however, it is
argued in sections~\ref{sec:formal} and \ref{sec:incoherence} of this
paper that the nucleons radiate incoherently and the energy flux is
insensitive to the exact baryon distribution.  Therefore, one can
approximate the true baryon density by a constant density inside the
nuclear radius, $\rho_{B}(r) \equiv \rho_{o} \Theta (R_{N}-r)$.

The motion of the baryons will be modeled as having a smooth
transition from their initial speed to their final speed. A Fermi-type
parameterization of the speed is used~\cite{MS95}:
\begin{equation}
  \beta(t) = \beta_{f} + \frac{\beta_{i} - \beta_{f}}{1 +
    \e^{t/\tau}},
\end{equation}
where $\tau$ is the stopping parameter.  Call the unit vector along
the beam axis $\vec{z}$.  Therefore, the first nucleus is traveling
with velocity $\beta \vec{z}$ and the second with $- \beta \vec{z}$.
The initial and final speeds are fixed from experimental results and
the stopping parameter is allowed to vary in this model.

The beam energy per nucleon is $E_{\rm beam}/A$.  The initial rapidity
is related to the beam energy by $M_{N} \cosh y_{i} = E_{\rm beam}/A$,
and the initial speed by $\beta_{i} = \tanh y_{i}$.  The energy lost
during the collision is characterized by the rapidity loss, $\delta y
= y_{i} - y_{f}$.  Mishustin \emph{et.\ al}~\cite{MS95} claim a
rapidity loss of $\delta y = 2.4 \pm 0.2$ for central Au+Au collisions
at RHIC energies of $E_{\rm beam}/A = 100$ GeV\@.  The same
value is used for this study of Pb+Pb at the same energy.  The
stopping parameter is allowed to vary between $5$ and $20$
fm.\footnote{Investigations up to 50 fm still maintain the
  characteristic shape of the angular energy flux distribution.}

\section{Incoherence}
\label{sec:incoherence}

The initial speed of the nucleons for $E_{\rm beam}/A = 100$ GeV is
$\beta_{i} \approx 0.99996$.  The velocity of the nucleons relative
to the center-of-mass is small compared to the velocity of the
center-of-mass.  This means that the collision time between nuclei
will be much shorter than the interaction time between nucleons in the
same nucleus.  The nuclei then can be treated as static collections of
nucleons with uniform longitudinal motion.  This is similar to the
frozen approximation used in the parton model.

The baryon current can be represented as a collection of point
particles:
\begin{equation}
  \vec{B}(\vec{x},t) = \vec{\beta}^{(1)}(t) \sum_{j=1}^{A} \delta[ \vec{x}
  - \vec{X}_{1}(t) - \vec{r}_{j}] + \vec{\beta}^{(2)}(t) \sum_{j=1}^{A}
  \delta[ \vec{x} - \vec{X}_{2}(t) - \vec{r}_{j}]
\end{equation}
where the first (second) term represents the first (second) nucleus,
$\vec{X}_{i}(t)$ is the position of the center of mass of the
$i^{\mathrm{th}}$ nucleus and $\vec{r}_{j}$ is the position of the
$j^{\mathrm{th}}$ baryon relative to the center of mass at
$t=-\infty$.  Since the lab frame coincides with the equal-velocity
frame, $\vec{X}_{1} = - \vec{X}_{2} \equiv \vec{X}$.  The position is
the integral of the velocity with the boundary condition that the
velocity asymptote to the free particle solution at $t = \pm \infty$.
The exact form used is
\begin{equation}
  \vec{X}(t) = \int^{t} \beta(t') \d t' \; \vec{z} = \beta_{i} t
  \vec{z}+ (\beta_{f} - \beta_{i})\tau \ln (1 + \e^{t/\tau})\vec{z}.
\end{equation}

Using this form of the current in equation~(\ref{freq}) gives
\begin{eqnarray}
  \label{freq2}
  \frac{\d ^{2} E}{\d \varepsilon \d \Omega} & = &
  \frac{\varepsilon^{2}g_{v}^2}{\pi^{2}} \; \mathcal{P} \int \d t \int
  \d t' \;[\beta(t) \sin \psi(t)] [\beta(t') \sin \psi(t')] \e^{\i
    \varepsilon (t-t')} \\ \nonumber & & \; \; \times \sin
  [\varepsilon \phi(t)] \sin [\varepsilon \phi(t')]
\end{eqnarray}
where $\sin \psi(t) = \vec{n} \times (\vec{n} \times \vec{z})$ at time
$t$, $\phi(t) = \vec{n} \cdot \vec{X}(t)$ and ${\mathcal P}$ is the
coherence factor defined in equation~(\ref{eq:P}).

If in the coherence factor, $|\varepsilon \, \vec{n} \cdot
(\vec{r}_{j} - \vec{r}_{k})| \ll 1$ then the phases are all near zero.
If, on the other hand, \mbox{$|\varepsilon \, \vec{n} \cdot
  (\vec{r}_{j} - \vec{r}_{k})| > 1$} then the phases are large and
nearly random.  In the end, when a statistical average over all
possible configurations is taken using the ground-state
wave-functions, the first case leads to $\mathcal{P} \approx A^{2}$
corresponding to the nucleons radiating as one large baryon of
``baryonic charge'' $A$.  For the latter case the coherence factor
becomes $\mathcal{P} \approx A$ corresponding to each nucleon
radiating independent of its neighbors.  Since neither of these
conditions hold absolutely, the coherence factor will be some power of
the atomic weight, $A^{\alpha}$ where \mbox{$1 \leq \alpha \leq 2$}.
To neglect the vector meson mass, its outgoing energy must be on the
order of several GeV\@.  The inter-particle distance is on the order
of 1 fm.  The latter condition holds and the nucleons should radiate
incoherently, $\alpha = 1$.

\section{Attenuation}
\label{sec:attenuation}

Throughout this work the vector meson field has been treated
analogously to the electromagnetic field with just a change in
coupling strength.  However, there is one aspect where there is a
significant difference.  The vector mesons interact strongly with the
baryons; therefore, there will be an attenuation of vector mesons as
they pass through either the nucleus from which they are radiated or
the other nucleus.  The attenuated energy flux can appear as radiation
through another channel or could be used to heat the nuclear material.
Since most of the radiation will occur early during the collision
[equation~(\ref{eq:angular}) is dominated by the denominator which
increases during the collision] and is directed in the forward
direction, this attenuation is non-negligible.  This reduction is
accounted for by including a multiplicative factor, ${\mathcal
  A}^{2}(t)$, in the radiated power,
\begin{equation}
  \frac{\d P(t')}{\d \Omega} = R^{2} [{\mathcal A}^{2}(t) \vec{S}(t)
  \cdot \vec{n}]_{\rm ret}.
\end{equation}
Forms of equations (\ref{eq:angular}) and (\ref{eq:totalE}) involving the
attenuation factor follow the development in section~\ref{sec:formal}.

This attenuation factor\footnote{It is assumed here that the vector
  meson will interact with the baryons immediately after it is
  created.  If the nucleus demonstrates color transparency, then this
  attenuation factor will be reduced.} is taken as,
\begin{equation}
\label{atten}
{\mathcal A}(t) = \frac{\int \d ^{3}x\, \rho_{B}(\vec{x}) \e^{-l
    \sigma \rho_{B}} \e^{-\chi(t) \gamma' \sigma \rho_{B}}} {\int \d
  ^{3}x\, \rho_{B}(\vec{x})} ,
\end{equation}
where $l$ and $\chi$ are the distances traveled through the radiating
nucleus and the second nucleus respectively, $\sigma$ is the vector
meson-nucleon total cross-section, and $\rho_{B}$ is the baryon
density.  Equation~(\ref{atten}) can be evaluated in the rest frame of
the radiating nucleus making $l$ time-independent.  To maintain the
covariance of the model the attenuation factor is evaluated along the
axis of motion and taken to be independent of outgoing angle.  Since
most of the radiation is in the far forward direction, this results in
only a small overestimation of the attenuation.

To evaluate equation~(\ref{atten}), consider the radiation coming from
the nucleus traveling in the positive $\vec{z}$ direction (the
argument is identical for the other nucleus).  Move the origin for the
integration to the center of mass in nucleus' rest frame.  The
distance traveled inside this nucleus depends only on where the
radiation originates (remember, the attenuation is assumed independent
of outgoing angle).  Using cylindrical coordinates $(\rho, \phi, z)$
to take advantage of the azimuthal symmetry, we find
\begin{equation}
  l = -z + \sqrt{R_{N}^{2} - \rho^{2}} ,
\end{equation}
where $R_{N}$ is the nuclear radius.  Since the radiation must come
from inside the nucleus the following condition \emph{always} holds:
$z^{2} + \rho^{2} \leq R_{N}^{2}$.  The form of the equation for the
distance through the target nucleus is more involved since the nucleus
is moving.  Defining the distance between the two centers of mass as
$d = - 2 \gamma'(t) X(t)$, the distance traveled through the moving
nucleus is
\begin{equation}
  \chi(t) = \left\{ \begin{array}{lll} \frac{2}{\gamma'} \sqrt{R^2 -
        \rho^{2}} & ; & \mbox{$d - R_{N}/\gamma' > z$} \nonumber \\ -
      (z-d) + \frac{1}{\gamma'} \sqrt{R_{N}^{2} - \rho^{2}}& ;
      \hspace{3mm} & \mbox{$d - R_{N}/\gamma' < z < d +
        R_{N}/\gamma'$} \nonumber \\ 0 & ; & \mbox{otherwise}
      \nonumber
      \end{array}  \right.
\end{equation}
where $\gamma'$ is the Lorentz factor for the target moving in the
rest frame of the radiating nucleus.  It is related to the Lorentz
factor in the lab frame [equation~(\ref{lorentz})] by
\begin{equation}
  \gamma'(t) = \gamma^{2}(t) (1 + \beta^{2}) = \frac{1+\beta^{2}}{1 -
    \beta^{2}}.
\end{equation}

\section{Energy Spectrum}
\label{sec:massless}

Throughout this paper the vector meson mass has been assumed small
compared to the particles' outgoing energy, and hence neglected.  It
is the purpose of this section to justify this assumption.
Equation~(\ref{freq2}) gives the energy distribution of the radiated
mesons.  If the observer is assumed far away, the angle to the
observer, $\psi(t)$, will change little over the time of interest.
The $\sin \psi$ factor may be taken outside the time integral as a
constant.  We define the meson energy spectrum by keeping only the
energy and time dependent factors in equation~(\ref{freq2}),
\begin{equation}
\label{en_spectrum}
{\mathcal I}^{2}(\varepsilon) = \left| \; \varepsilon
  \int_{-\infty}^{\infty} \d t \; \vec{\beta}(t) \e^{\i \varepsilon t}
  \sin [\varepsilon \vec{n} \cdot \vec{X}(t)] \; \right|^{2}.
\end{equation}
Note that ${\mathcal I}^{2}(\varepsilon)$ is dimensionless.  Again
assuming the angle to the observer changes little and the velocity
remains near one, the integral can be evaluated analytically giving:
\begin{eqnarray}
  {\mathcal I}^{2}(\varepsilon) & = & (\varepsilon \tau)^{2} \left|
    \rule{0mm}{5mm} \; B[\i \varepsilon \tau (1 - \beta_{i} \cos
    \psi), \i \varepsilon \tau (\beta_{f} \cos \psi - 1)] \right.
  \nonumber \\ & & + \left. B[\i \varepsilon \tau (1 + \beta_{i} \cos
    \psi), -\i \varepsilon \tau (\beta_{f} \cos \psi + 1)] \;
    \rule{0mm}{5mm} \right|^{2},
\end{eqnarray}
where $B[x,y]$ is the beta function which is related to the gamma or
factorial function by
\begin{equation}
  B[x,y] \equiv \frac{\Gamma(x) \Gamma(y)}{\Gamma(x+y)}.
\end{equation}

\begin{figure}
  \centering\epsfig{figure=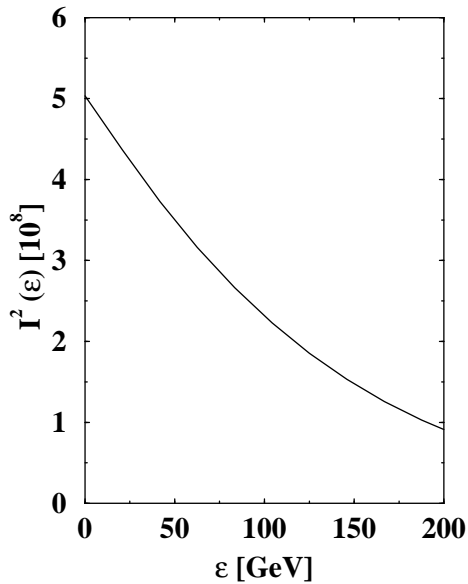,width=.46\linewidth}
\caption{The dimensionless energy spectrum, ${\mathcal I}^{2}
  (\varepsilon)$, as defined in equation~\protect\ref{en_spectrum} as
  a function of outgoing vector meson energy.}
\label{fig:spectrum}
\end{figure}

As can be seen in figure~\ref{fig:spectrum} the peak occurs at
$\varepsilon = 0$, and decreases slowly as $\varepsilon$ increases.
This slow decrease means that most of the radiation is above the
vector meson mass and that the massless assumption is acceptable.
Cutting the spectrum off at the meson mass $\varepsilon = m_{\omega} =
.783$ GeV, reduces the integrated spectrum by less than $1\%$.

\section{Results}
\label{sec:results}

Throughout this work, only central collisions of identical nuclei at a
bombarding energy achievable at RHIC of $E_{\rm beam}/A =
100$~GeV~\cite{Oz91} have been considered.  A short discussion is
still needed concerning the value of the required coupling constant,
$g_{v}$.  There is no solid evidence to direct a choice of the
coupling constant for the type of reaction that this work describes.
A value that reproduces static nuclear properties has been
chosen~\cite{W95}; $g_{v}^{2}/4\pi = 10.8$.  Using a value of
$g_{v}^{2}/4\pi$ between eight and nine, Gross, Van Orden and
Holinde~\cite{Gr92} were able to reproduce free nucleon-nucleon phase
shifts to several hundred MeV\@.  Since the $(\sigma, \omega)$ model
is not an asymptotically-free theory, the coupling constant may even
increase at higher energy and momentum transfer; this would lead to
even more radiation than is shown below.  The paper by Mishustin
\textit{et.\ al}~\cite{MS95}, uses a coupling constant of
$g_{v}^{2}/4\pi = 15.1$, stronger than what is used here.  As long as
the coupling constant is of at least order 1, this model predicts
appreciable amounts of energy radiated through this channel.  This is
a crude model that has incorporated several simplifying assumptions.
It would be inconsistent to think of the predicted numbers here as
more than a rough guide to the order of magnitude of what may actually
be seen.  To remove any ambiguity in the coupling constant, the
angular distributions are shown with the coupling constant divided
out.

\begin{figure}
\noindent
\begin{minifig}[t]
  \centering\epsfig{figure=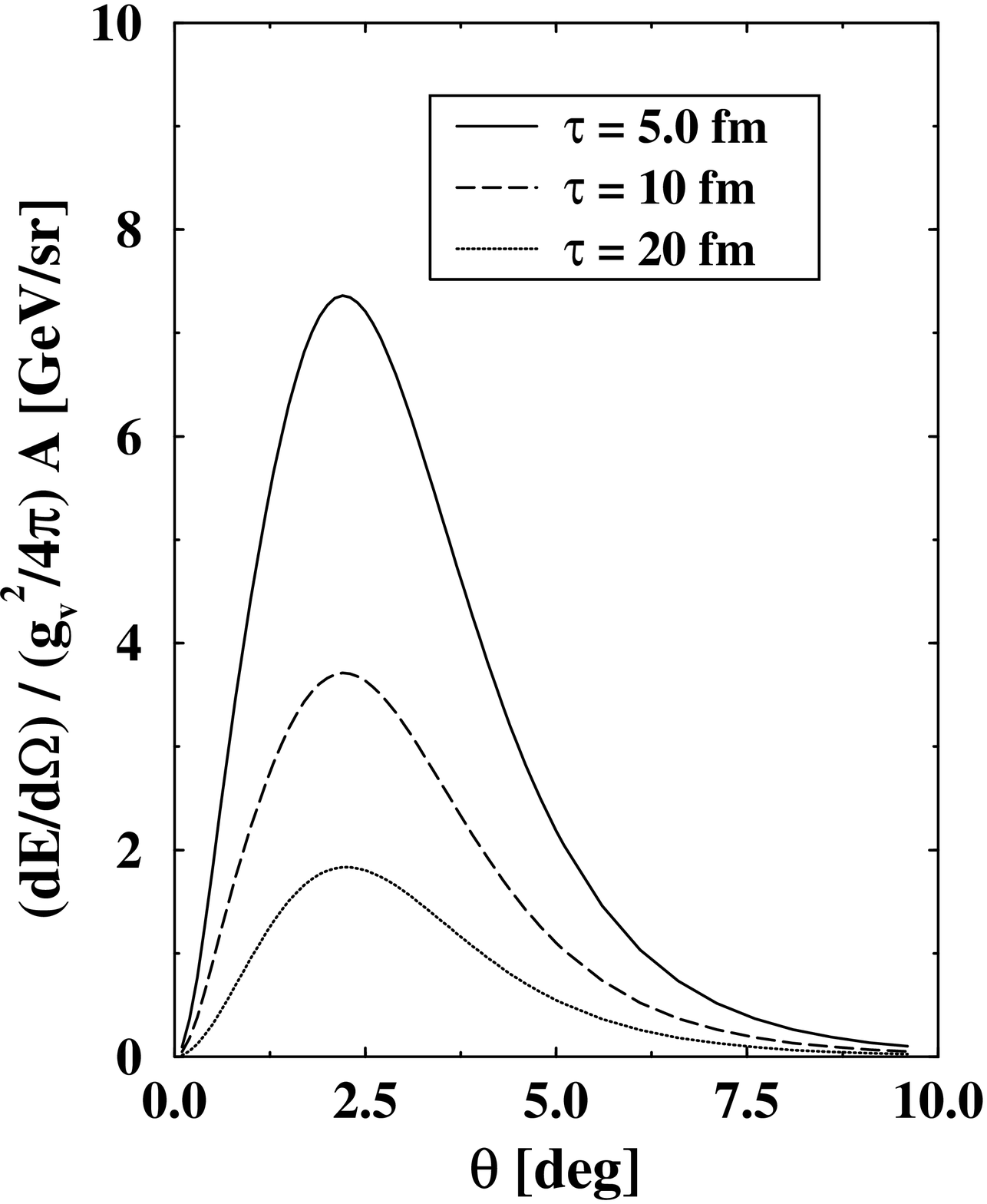,width=\linewidth}
\caption{Angular distribution of energy flux as a function of lab
  angle for outgoing vector mesons.  The rapidity loss is fixed at
  $\delta y = 2.4$ and the vector meson-nucleon total cross-section is
  fixed at $\sigma = 30$ mb.  It is shown for $\tau = 5.0$ fm (solid
  line), $\tau = 10$ fm (dashed line) and $\tau = 20$ fm (dotted
  line).  The $A$ and $g_v^2/4\pi$ factors have been divided out.}
    \label{fig:stopping}
\end{minifig}\hfill
\begin{minifig}[t]
  \centering\epsfig{figure=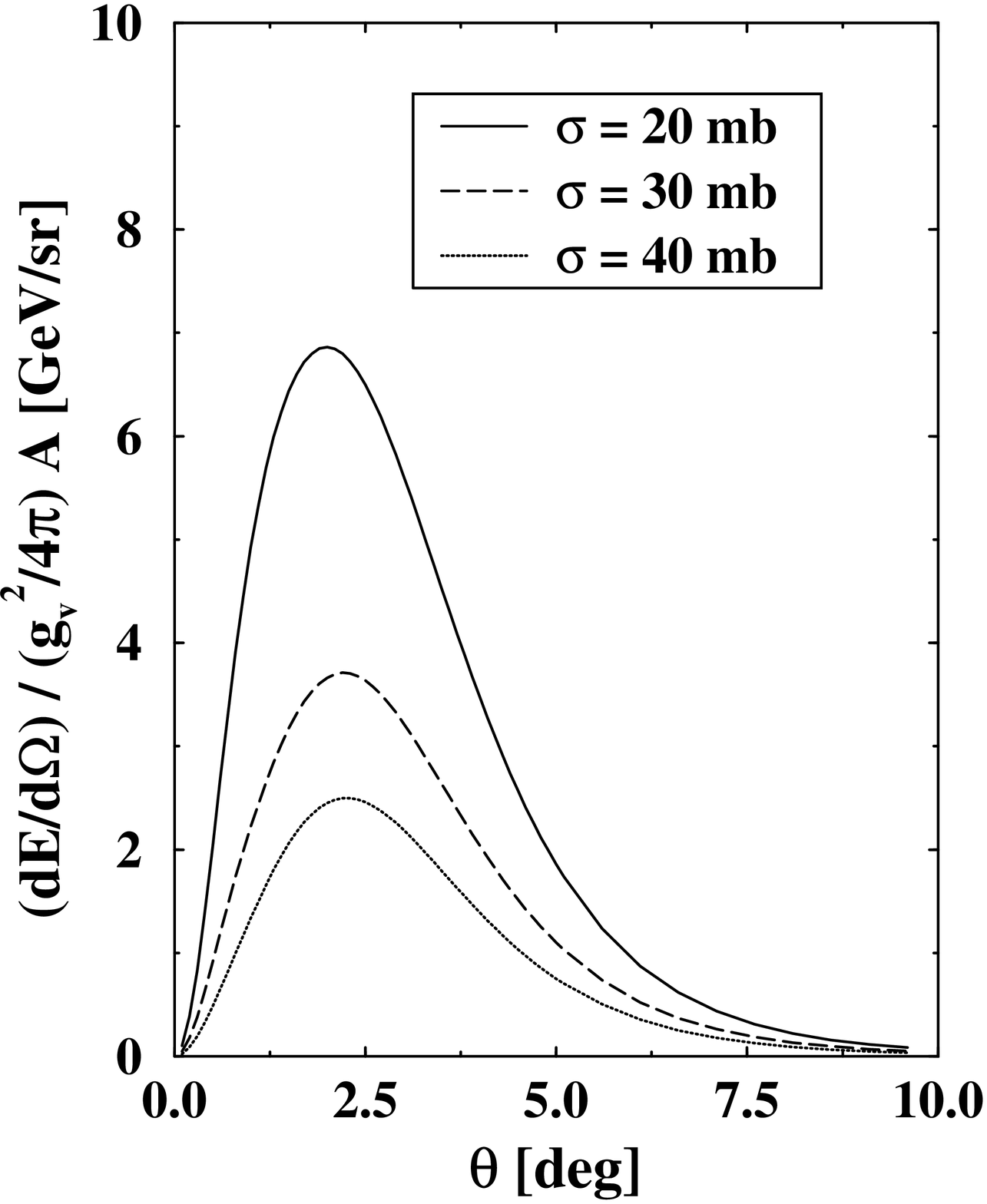,width=\linewidth}
\caption{Same as figure \protect \ref{fig:stopping}
  except stopping parameter is fixed at $\tau = 10$ fm and the
  cross-section takes the values $\sigma = 20$ mb (solid line),
  $\sigma = 30$ mb (dashed line) and $\sigma = 40$ mb (dotted line).}
    \label{fig:cross}
\end{minifig}
\end{figure}

\begin{figure}
  \centering\epsfig{figure=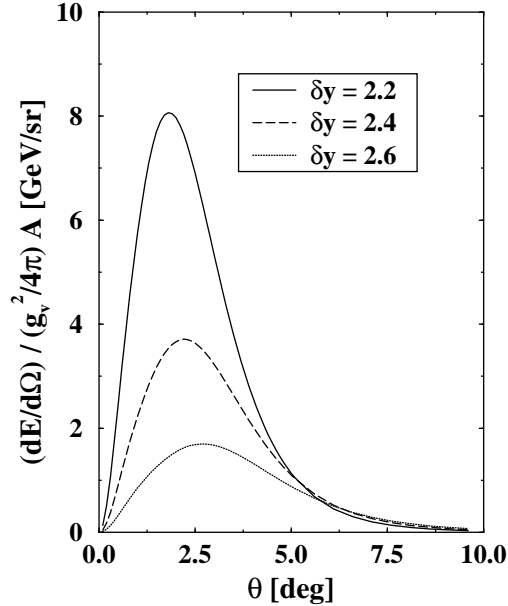,width=.46\linewidth}
\caption{Same as figure \protect \ref{fig:stopping}
  except stopping parameter is fixed at $\tau = 10$ fm and the
  rapidity loss takes the values $\delta y= 2.2$ (solid line), $\delta
  y = 2.4$ (dashed line) and $\delta y = 2.6$ (dotted line).}
    \label{fig:rapidity}
\end{figure}

Figures \ref{fig:stopping}, \ref{fig:cross} and \ref{fig:rapidity}
show the angular distribution of radiated energy in the form of vector
mesons.  Following the analysis of section~\ref{sec:incoherence}, the
energy flux is calculated as the incoherent sum of $2A$ nucleons
radiating as point particles,
\begin{equation}
  \frac{\d E}{\d \Omega} = \frac{g_v^2}{4\pi} \; A
  \int_{\infty}^{\infty} \d t \left( \frac{{\mathcal A}^2(t)
      \dot{\beta}^{2}(t) \sin^{2} \theta}{(1 - \beta(t) \cos
      \theta)^{5}} + \frac{{\mathcal A}^2(t) \dot{\beta}^{2}(t)
      \sin^{2} \theta}{(1 + \beta(t) \cos \theta)^{5}} \right) .
\end{equation}  
The factor $(g_v^2 / 4\pi) A$ is divided out in the figures to remove
any ambiguity.\footnote{The dimensions of the remaining expression are
  converted by 1 fm$^{-1}$ = 0.197 GeV.}

In each of the figures, two of the parameters are held constant while
the third is varied over reasonable values.  The peak of energy flux
occurs around $2.5^{o}$ off the beam-axis.  It is seen that
variations in the stopping time, $\tau$, and the cross-section,
$\sigma$, only affect the magnitude and not the characteristic angular
distribution shape.  Changes in the rapidity loss, $\delta y$, affect
both the magnitude and shape of the distribution with larger rapidity
losses in this range causing less radiation and causing the peak to be
moved further off axis and broadened.  It is interesting to note that
the smallest rapidity loss used in this study, $\delta y = 2.2$,
results in the most energy being radiated.  This can be explained by
considering the circumstances under which most of the energy is
radiated.  At relativistic speeds, the energy radiated is dominated by
the factor of $(1-\vec{n} \cdot \vec{\beta})^{-5}$ in
equation~(\ref{eq:angular}).  A smaller rapidity loss results in the
nucleus spending more time traveling faster, resulting in more
radiation.  Of course this only holds over a small region.  At some
point the smallness of the acceleration overcomes the smallness of the
denominator and at zero rapidity lost, there is no radiation.

For the intermediate values of the parameters ($\tau = 10$ fm, $\sigma
= 30$ mb and $\delta y = 2.4$) and $A = 208$ (Pb beams) the total
energy radiated, $E_{\rm rad}$, is $353$ GeV\@.  The total energy
available to be radiated in this collision is $3.78 \times 10^{4}$
GeV\@.  Reducing the stopping parameter to $\tau = 5$ fm gives
$E_{{\rm rad}} = 730$ GeV\@.  A rapidity loss of $\delta y = 2.2$ and
a stopping parameter of $\tau = 10$ fm gives $E_{{\rm rad}} = 516$ GeV
out of a total available energy of $3.70 \times 10^{4}$ GeV\@.  On the
order of 1\% of the total energy loss is through this channel.

\section{Conclusions}
\label{sec:conclusions}

In this paper a model for treating the bremsstrahlung radiation of
neutral vector mesons coupled to a baryon current during central
relativistic heavy-ion collisions has been developed.  This model
treats the nuclei as clusters of baryons frozen in relative position
over the time of the collision.  The clusters' velocities are modeled
as changing smoothly with time.  Lorentz contraction greatly increases
the baryon density relative to the scalar density; hence, the scalar
meson radiation becomes negligible at high energies.  Most of the
energy is radiated as highly energetic vector mesons, allowing for
their mass to be neglected.  The vector mesons interact strongly with
the baryons and therefore, an attenuation factor is included.  To
maintain the covariance of the model, this attenuation factor is
assumed independent of angle.

The modeling of the flow of the baryon current is the only freedom in
the model.  By using a smooth connection between the initial rapidity
of the nuclei with the experimentally measurable final rapidity, the
baryon current depends on just two parameters, the rapidity loss
(obtained from the initial and final rapidities) and a stopping
parameter which is allowed to vary freely.  Although this model is
used well outside its tested domain, in the end only the vector field
coupling to the conserved baryon current is seen.  The model predicts
a characteristic shape for the angular distribution of radiation that
is robust against parameter variations.  It appears this mechanism may
contribute to meson production in the next generation of relativistic
heavy-ion collisions like those at RHIC\@.  This possibility would
seem to be worth exploring experimentally.\footnote{If this process
  were to be observed, it could provide a diagnostic for the baryon
  flow during such collisions.  For example, only straight line motion
  of the baryons has been considered.  The radiation is focused
  sharply around the instantaneous motion of the baryon current so
  deviations from the angular distribution shown here could be used to
  track the intermediate flow of baryons during the collision.  See
  also Ref.~\cite{I89}.}

\begin{ack}
  The author wishes to thank J. Dirk Walecka for many
  thought-provoking discussions and comments throughout this work and
  S. Pratt for a helpful observation.  This work was supported in part
  by the Department of Energy grant No. DEFG0594ER40829 and a
  SURA/CEBAF fellowship.
\end{ack}

\end{document}